\documentclass[aps,prla,twocolumn,superscriptaddress]{revtex4-1}

\usepackage{graphicx}
\usepackage{xcolor}

\begin{document}

\title{Coherent Cerenkov radiation and laser oscillation in a photonic crystal}

\author{T.~Denis}
\author{M.W.~van~Dijk}
\author{J.H.H.~Lee}
\author{R.~van~der~Meer}
\author{A.~Strooisma}
\author{P.J.M.~van~der~Slot} \email{p.j.m.vanderslot@utwente.nl}
\affiliation{Laser Physics and Nonlinear Optics (LPNO),\\ Department of Science and Technology and MESA\textsuperscript{+} Institute for Nanotechnology, \\ University of Twente, P.O.Box 217, 7500 AE Enschede, The Netherlands}
\author{W.L.~Vos}
\affiliation{Complex Photonics Systems (COPS), \\ Department of Science and Technology and MESA\textsuperscript{+} Institute for
Nanotechnology, \\ University of Twente, P.O.Box 217, 7500 AE Enschede, The Netherlands}
\author{K.-J.~Boller}
\affiliation{Laser Physics and Nonlinear Optics (LPNO),\\ Department of Science and Technology and MESA\textsuperscript{+} Institute for Nanotechnology, \\ University of Twente, P.O.Box 217, 7500 AE Enschede, The Netherlands}

\date{\today}


\begin{abstract} We demonstrate that photonic crystals can be used to generate powerful and highly coherent Cerenkov radiation that is excited by the injection of a beam of free electrons. Using theoretical and numerical investigations we present the startup dynamics and coherence properties of such laser, in which gain is provided by matching the optical phase velocity in the photonic crystal to the velocity of the electron beam. The operating frequency can be varied by changing the electron beam energy and scaled to different ranges by varying the lattice constant of the photonic crystal.
\end{abstract}

\maketitle
\section{Introduction}
Controlling the fundamental strength of light and matter interaction with nano photonic structures is of fundamental importance for the generation of radiation, such as demonstrated with parametric emission, \cite{Krauss_2009} in coherent interaction with vacuum fluctuations \cite{Deppe_2004} and with controlling spontaneous emission \cite{Lodahl2004, Englund_Yamamoto_PRL_2005}. Also stimulated emission, \textit{i.e.}, the amplification of light, has been enhanced using nano photonic structures. Prominent examples are nanolasers employing point and line defect cavities in two-dimensional \cite{Painter1999, Kim_Lee_Science_2004, Noda_Review_Science_2006, Noda_Nat_2013, Yacomotti_NatPhot_2015} and three-dimensional photonic crystals \cite{Tandeachanurat_Arakawa_NatPhot_2011}. Bloch mode lasers operating near the edge of the Brillouin zone  have been realized in two-dimensional \cite{MeierAPL_1999, ImadaNoda_APL_1999, NotomiAPL_2001, NodaScience_2001, RyuAPL_2003} and three-dimensional photonic crystals \cite{Cao_NatMat_2002}. In cavity-based nanolasers the main function of the photonic crystal with its periodic variation of the dielectric constant at the scale of the wavelength \cite{Yablonovitch1987,Joannopoulos2008} is providing strong feedback for field enhancement in a small, wavelength-scale mode volume. In Bloch mode lasers the photonic crystal provides field enhancement via a reduced group velocity and forms a distributed feedback laser, which offers larger mode volumes and output. However, in all these photonic crystal lasers, the amplification of light is provided by conventional gain media, specifically, semiconductor quantum wells, quantum dots, or organic dyes. This principally limits the laser output wavelengths to the bound-electron transitions of the respective gain material. 

Here we demonstrate a very different and much more general approach to photonic crystal Bloch mode lasers where gain and coherent output radiation is provided by free electrons, without relying on any specific gain material. Thereby, the range of output wavelengths is not bound to pre-determined values but can be scaled over orders of magnitude via scaling the spatial period of the photonic crystal \cite{vanderSlot2012}. The work that we present here gives the first complete description and proof of coherent emission from a photonic crystal-based free-electron laser, by solving self-consistently the coupled Maxwell and Lorentz-Newton equations in three dimensions. Thereby we obtain access to the full nonlinearity in the dynamics of the field, such as startup from noise leading into steady-state oscillation, followed by mode competition and spectral condensation, and to the dynamics of the electron beam, such as space-charge effects and gain saturation. Using numerical modeling we show that a beam of electrons in a photonic crystal can generate laser radiation with significant power and with high spectral and spatial coherence. In our analysis we present an example with continuous-wave output in the kW-range, emitted into a single spatial mode at a single frequency. The key for obtaining such high-brightness radiation is a proper choice of the photonic crystal parameters as to maximize the mutual feedback between the crystal-internal radiation and the electrons. Thereby, the radiation process is brought into the regime of amplification of light, \textit{i.e.}, stimulated emission occurs, leading to laser oscillation and spectral condensation with high coherence.

The basic principle of light generation by free electrons in photonic crystals can be understood in terms of the Cerenkov effect \cite{Cerenkov1934,Cerenkov1937,Frank1937} where the photonic structure strongly modifies the dispersion relation. Calculations have shown that the spectral distribution and emission pattern can be varied over wide ranges \cite{Luo2003,Kremers2009a} via the period of the photonic crystal. Scaling the output to desired wavelengths is of high interest for providing novel light sources, e.g., in the extreme ultraviolet (EUV) radiation \cite{Andre2003}. However, all these calculations and scaling considerations are based on single, \textit{i.e.}, non-collectively interacting, point charges traveling with a constant velocity. Thereby this describes only the generation of spontaneous emission with low efficiency and with low spectral and spatial coherence. 
So far there is no study of stimulated emission or laser oscillation driven by free electrons in photonic crystals. Experimental investigations are restricted to one-dimensionally periodic Bragg structures and spontaneous emission which generates incoherent radiation at low efficiency. For instance, using alternating stacks of thin films, soft x-ray radiation has been observed with relativistic electrons \cite{Yamada1999}. Also non-relativistic electrons, with energies of about $30\,\mathrm{keV}$, were used in multilayer stacks. In the latter, the electrons were sent through a hole in the sample and generated visible and near infrared radiation \cite{Adamo2009}. In all cases, the observed radiation was temporally incoherent (spectrally broadband) and spatially multi-mode.  The generation efficiency was low such that highly sensitive detection was required, \textit{e.g.}, cryogenic detectors in the near infrared \cite{Adamo2009} or photon counting in the soft x-ray range \cite{Yamada1999}.

\begin{figure}[bt]
	\includegraphics[width=0.90\linewidth]{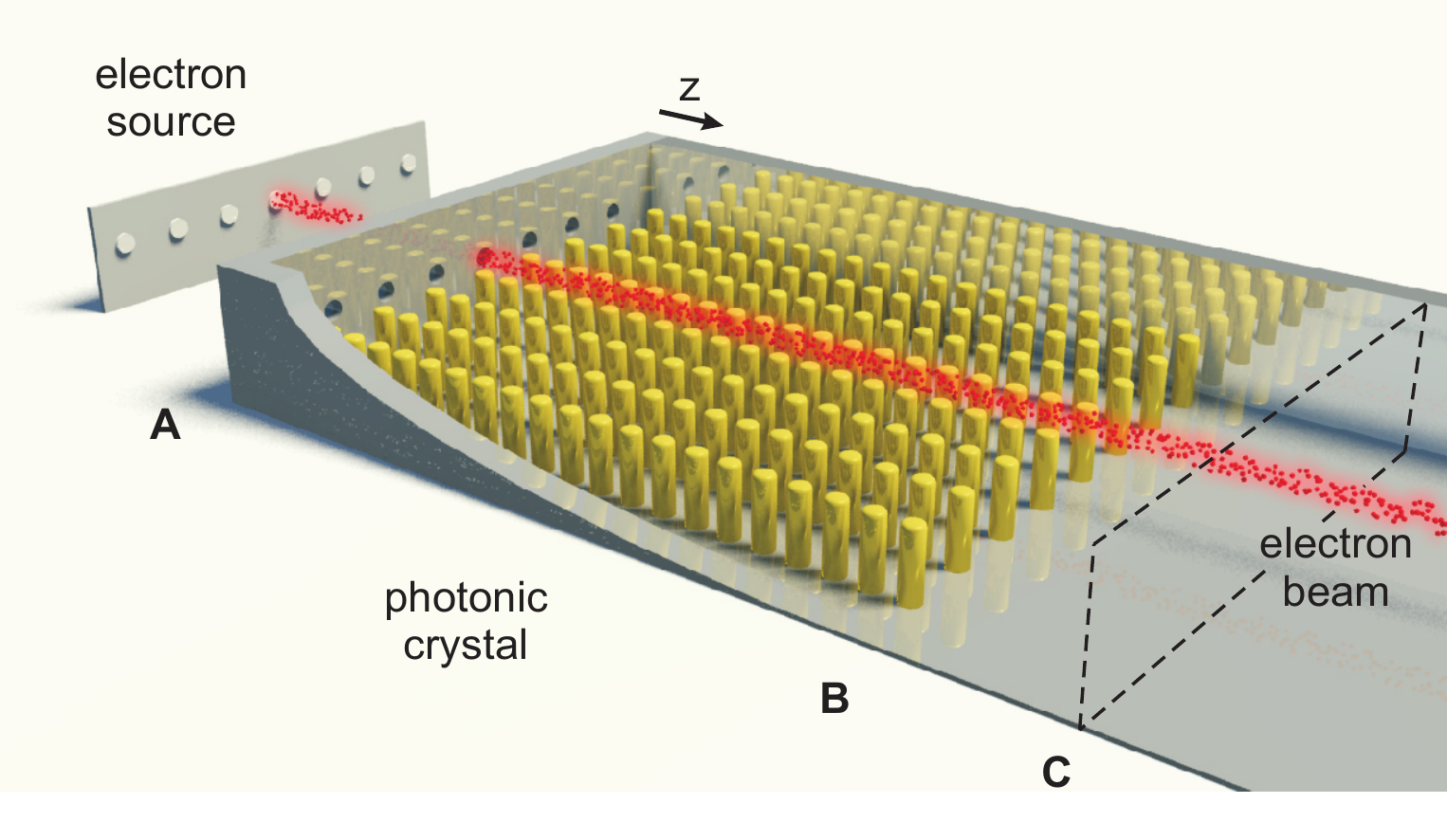}
	\caption{Generating coherent Cerenkov radiation in a photonic crystal (solid rods shown in yellow) embedded in a hollow waveguide. In the figure the upper wall and one of the side walls of the waveguide are removed to reveal the photonic crystal and electron beam. A continuous electron beam (indicated in red) enters through plane A. The end of the photonic crystal at plane B functions as a partial reflector for the radiation. The transmitted radiation is analyzed at plane C.
	\label{figure1}
}
\end{figure}

\section{Cerenkov lasing in a photonic crystal}
To explain our approach, we refer to Fig.~\ref{figure1} which shows a continuous beam of electrons propagating through a photonic crystal slab. The slab consists of a two-dimensional photonic crystal made of an array of solid rods contained in a hollow waveguide. The electron beam enters in plane A and leaves, together with the generated electromagnetic waves, through plane C.

Let us first consider the simplest case of a single electron that propagates through the crystal  along a given trajectory with constant velocity as described in Refs.~\cite{Luo2003,Kremers2009a} thereby neglecting any back action of generated radiation on the electron trajectory. Referring to Fig.~\ref{figure1}, the electron induces a transient (pulsed) polarization in each rod that it passes. The polarization acts as the source for a radiation pulse which spreads throughout the photonic crystal through multiple scattering. This generation and scattering process repeats with each passage of the electron along the next rod. Via linear superposition, the phasing of the single-rod responses determines the electromagnetic field pattern that builds up in the crystal. This pattern generally comprises a large manifold of excited Bloch modes, \textit{i.e.}, it will exhibit a complicated spatial structure, contain a wide range of optical frequencies, and show a complex temporal dynamics. Nevertheless, the single-electron radiative response of the photonic crystal is fully deterministic as it is given by the electron trajectory.

Next we consider that a constant-current beam of mono-energetic electrons is injected along the same trajectory, still neglecting radiation feedback and Coulomb repulsion. When considering a particular  frequency to be generated or a spatial mode to be excited, all electrons contribute with the same absolute value in field amplitude. In real electron beams, however, the beam current is only approximately constant in space and time due to a small noise component, because the positions of the electrons are randomly distributed along the beam. The latter is due to thermal noise in the beam and also due to the quantum (wave) nature of the electrons \cite{Benson85}. The result is a random phasing of the single-electron field contributions, leading to destructive interference in emission, except for a small shot-noise contribution. Due to its origin from noise, the emission is weak, randomly phased, and fluctuates in power, \textit{i.e.}, the output is incoherent.  

In sharp difference to incoherent emission where the radiation feedback is negligible, in the approach presented here we maximize the feedback in order to enhance the mutual interaction between electrons and the electromagnetic field. This feedback leads to phasing of the single-electron responses and imposes coherence in emission. To describe how feedback is maximized, we recall that for a given light frequency, $\nu$, the Bloch eigenmodes describing the spatially periodic distribution of the electric field inside a photonic crystal can be decomposed into spatial harmonics of order $\mathrm{m}$ with Fourier amplitudes \textbf{E}$^m$ \cite{Joannopoulos2008}. In this decomposition, the lowest-order Fourier components with $\mathrm{m}$~=~0 (1, 2, ...),  \textit{i.e.}, having a wave vector in the first (2nd, 3rd...) Brillouin zone, usually possess the highest field amplitudes. For maximizing feedback at a given electron beam velocity and a given frequency, as will be explained with Fig.~\ref{figure2} for a particular example, we chose the photonic crystal period such that the phase velocity of a low-order spatial harmonic becomes appropriately slowed in order to match it to the velocity of the electron beam. Provided that the electron beam propagates into the longitudinal direction ($\mathrm{z}$~-~direction) it is essential to select for the velocity matching a Bloch mode that possesses a longitudinal electric field component ($\mathrm{z}$~-~component $E^m_z$), because only then the electric field can reduce the kinetic energy of the electrons (deceleration) by increasing the field energy (amplification). Finally, to resonantly enhance the electric field we select the velocity matching to occur at a small group velocity, \textit{i.e.}, where the dispersion curve displays a small slope.  

The physical effects of these choices are conveniently discussed by considering the lowest order, velocity-matched spatial harmonic of the radiation field. When the beam is mono-energetic with a constant beam current, all beam electrons are initially at rest  relative to this spatial harmonic field component and provide a homogeneous charge density along the $z$-direction, except for the small shot-noise component mentioned above. The longitudinal electric field component, \textit{$E^m_z(z)$}, is therefore initially stationary with respect to the electrons but the direction of \textit{$E^m_z$} varies sinusoidally with $z$. After some interaction time, the feedback of the sinusoidal field on the electron beam, via position-dependent acceleration and deceleration, leads to the formation of electron bunches with a spatial periodicity and at coordinates that are given by the initial Cerenkov emission wavelength and phase at the velocity matched frequency. The bunching temporally synchronizes  the radiative response of the electrons into a collective response, such that their contributions add up coherently at the velocity matched harmonic. The described dynamics of the electrons, induced by radiation feedback, is actually well-known as the mechanism that provides stimulated emission (amplification of radiation) in free-electron lasers \cite{Elias1976} and other free-electron based coherent radiation sources, such as magnetrons and travelling wave tubes \cite{Gilmour2011}. In a free-electron laser, electron bunching is obtained when the phase velocity of the so-called ponderomotive force matches the electron velocity, which requires relativistic ($\gamma>>1$) electrons. To enable electron bunching when non-relativistic electrons ($\gamma\simeq 1$) are used, wave circuits, such as a helix in case of a travelling wave tube, are required to slow the longitudinal wave velocity to the beam velocity \cite{Gilmour2011}. Scaling to higher frequencies requires smaller wave circuits, which results in a strongly reduced output power due to higher circuit loss and lower current that can be transmitted through the device. In contrast, using a photonic crystal as wave circuit enables distribution of the electrons over many beams (see Fig. \ref{figure1}). By extending the transverse dimensions of the scaled photonic crystal the total current can be kept the same and even be increased, without the need to increase the current density and, consequently, conserve beam quality. Therefore, compared to other circuit-based sources, the reduction in output power should be significantly less if not absent when scaled for operation at higher frequencies.

\section{Methods}
A theoretical analysis of stimulated emission and laser oscillation driven by an electron beam in a photonic crystal requires self-consistently solving Maxwell's equations coupled to the Newton-Lorentz equations while imposing boundary conditions that appropriately describe the electromagnetic field in a photonic crystal. A self-consistent solution of these equations is not known so far. On the one hand, the radiation generated by a single point charge travelling with constant velocity has been calculated \cite{Luo2003,Kremers2009a,Yamada1999}. These approaches exclude radiation feedback and, thereby, do not include any amplification of radiation. Nevertheless, these calculations demonstrate the rich, multi-modal properties of Cerenkov radiation in photonic crystals, as well as the free scalability of the radiation wavelengths and frequencies with the crystal structure. On the other hand, models for free-electron lasers successfully describe the full, nonlinear dynamics within an electron beam that provides stimulated emission and gain saturation \cite{Milton2001, Emma2010}. However, the very different properties of the electromagnetic field inside a photonic crystal make it difficult to apply these models for describing stimulated emission, gain saturation and laser oscillation in a photonic crystal.

Our approach is based on a particle-in-cell numerical model \cite{CST, Burau_2010}, which self-consistently solves the  relativistic Newton-Lorentz equation,
\begin{equation}
\frac{d\gamma_{i} m_{i}\vec{v}_{i}}{dt}=q_{i}(\vec{E} + \vec{v}_{i}\times\vec{B}),  i=1,2,\cdots,N
\end{equation}
for each of the $N$ macroparticles, together with Maxwell's equations,
\begin{eqnarray}
  \nabla\cdot\varepsilon\vec{E}  &=& \rho \label{eq:m1} \\
  \nabla\cdot\vec{B} &=& 0 \label{eq:m2}\\
  \nabla\times\vec{E} &=& -\frac{\partial \vec{B}}{\partial t} \label{eq:m3} \\
  \nabla\times\frac{1}{\mu}\vec{B} &=& \frac{\partial\varepsilon\vec{E}}{\partial t} + \vec{J} \label{eq:m4}. 
\end{eqnarray}
Here, $\gamma_{i}=\left(1-\frac{v_{i}^{2}}{c^{2}}\right)^{-\frac{1}{2}}$ is the relativistic factor, $v_{i}=\left|\vec{v}_{i}\right|$, $\vec{v}_{i}=\frac{d\vec{r}_{i}}{dt}$ is the velocity of a macroparticle, $c$ is the speed of light in vacuum, $q_{i}$ and $m_{i}$ are the  charge and rest mass, respectively, of a marcoparticle, $\vec{E}$ and $\vec{B}$ are the total electric and magnetic fields, respectively, that includes the radiation fields as well as the self fields of the electrons and any external applied static magnetic field, and $\varepsilon$ and $\mu$ are the permittivity and permeability, respectively. The radiation and self fields are driven by $J$ and $\rho$, the current and charge densities, respectively, which are calculated from the position and velocity of the macroparticles. We note that eqs. \ref{eq:m1} to \ref{eq:m4} imply charge conservation: $\nabla\cdot\vec{J}=-\frac{\partial \rho}{\partial t}$. The radiation and self fields, which naturally emerge from Maxwell's equations, are subject to the boundary conditions imposed by the metallic photonic crystal slab, i.e., zero tangential component for the electric field and zero normal component for the magnetic field at the metal interfaces. With this approach we include all relevant radiation mechanisms (Cerenkov and transition radiation), the self-fields of the electron beam, the complete spectrum of Bloch eigenmodes and directions of the electric field as supported by the photonic crystal (propagating and evanescent). Our approach maintains the full nonlinearity in the dynamics in the field and the electron beam such as gain saturation, mode competition, spectral condensation and space-charge (Coulomb repulsion) effects. Using this method, we give for the first time a complete description of stimulated Cerenkov radiation and laser oscillation in a photonic crystal.

For a quantitative numerical modeling, we have chosen a finite-sized photonic crystal slab as shown in Fig.~\ref{figure1}. For investigating laser oscillation in the optical range, a dielectric material with $\mu$m~-~scale periodicity would have to be selected. However, in order to restrict the calculation times and the amount of data to what is feasible with our facilities, we chose a non-magnetic conducting material, for both the photonic crystal slab and the waveguide, and much longer periods in the mm~-~range, although this scales the generated frequencies down into the range of 10~GHz.  The photonic structure is based on a rectangular lattice with a longitudinal period $a_z$ and a transverse period  $a_x = 1.68a_z$. The lattice carries 20 x 7 cylindrical metallic rods (height $p = 1.6a_z$, diameter $d = 0.6a_z$) and is enclosed in a rectangular metallic waveguide (height  $h = 3.2a_z$ and width $w=13.44a_z$).  We assume all surfaces to be fully conducting, which is well-justified in the range of mm-waves as can be seen, e.g., from spatial mapping of individual field components inside photonic crystals \cite{Denis2012}. Consequently, the permeability and permittivity in eqs. \ref{eq:m1} and \ref{eq:m4} are set to the respective values for vacuum. For imposing single-sided output, a highly reflective surface is placed in plane A (Fig.~\ref{figure1}), which feeds-back radiation into the photonic crystal. The properties of radiation that has left the photonic crystal region in the \textit{z}-direction are analyzed in plane C, where the output field is decomposed into TE and TM modes of the empty waveguide, and where perfectly absorbing boundary conditions are applied. The walls of the metallic waveguide form the remaining boundaries of the simulation domain with zero tangential electric field and zero normal magnetic field as boundary conditions. To present at first the essential laser dynamics as discussed above, we begin with entering an ideal electron beam having zero velocity spread and emittance. The beam enters through a small aperture centered in plane A with a velocity of $0.23c$. We assume a stream of macroparticles that constitute an experimentally feasible constant current of $I = 1~\mathrm{A}$ \cite{Booske2008} with a beam radius of $1\,\mathrm{mm}$ and a rise time of 0.3 ns. The beam is guided along the \textit{z}-direction with a homogeneous and static magnetic field of $0.5~\mathrm{T}$.

\section{Results and discussion}
\begin{figure}[tb]
	\includegraphics[width=0.60\linewidth]{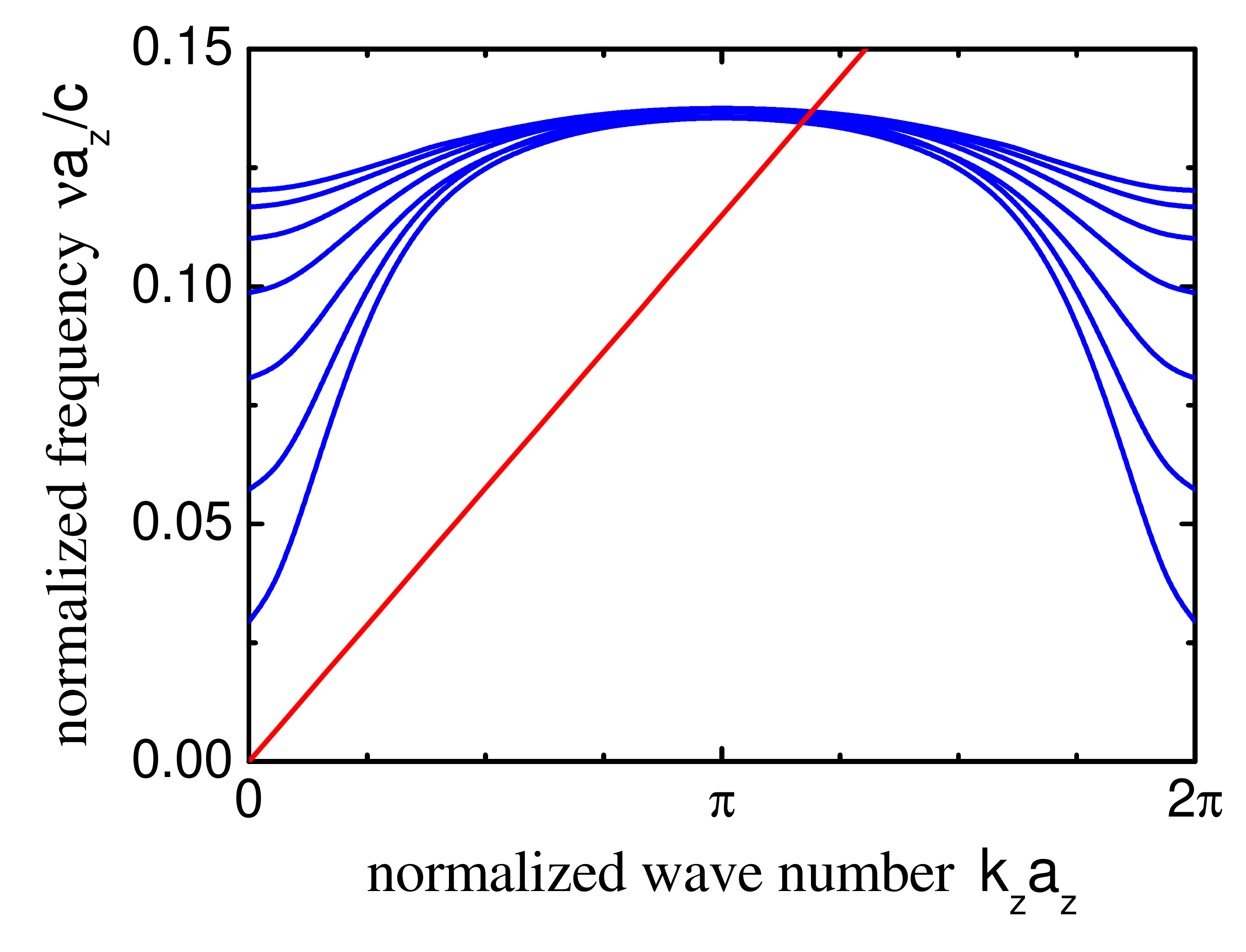}
	\caption{Dispersion of the lowest-frequency Bloch modes with a strong longitudinal electric field component. The straight line represents an electron velocity of $0.23c$. An intersection with a dispersion curve indicates the light frequency and wave vector at which the phase velocity is matching the electron velocity.
        \label{figure2}}
\end{figure}

The laser output frequency expected from velocity matching can be obtained by comparing the electron beam velocity with the electromagnetic phase velocity. The latter was calculated for the described metallic photonic crystal slab using an eigenmode method \cite{CST}. The dispersion diagram in Fig.~\ref{figure2} shows the lowest-frequency Bloch modes that possess an appreciable longitudinal field component, $E^m_z$, along the electron beam. The  straight line with a slope of $v_z= 0.23c$, $v_z$ being the longitudinal electron velocity, represents the electron beam dispersion  \cite{Gould1959}. Velocity matching is present at the intersection of the straight line with the dispersion curves of the Bloch modes. It can be seen that this occurs near a frequency of $\nu=0.13*a_z/c$, where the group velocity is small, about $c/20$. Note that for this example of velocity matching the group velocity is opposite to the phase velocity.

For definiteness we chose a crystal period  of $a_z = 2.5\,\mathrm{mm}$. According to Fig.~\ref{figure2} this  imposes velocity matching at frequencies around $16~\mathrm{GHz}$ where the assumption of perfectly conducting surfaces is well-justified. Furthermore, to ease a future experimental demonstration, with these periods and frequencies the fabrication of photonic crystals and a direct detection of electromagnetic fields is straightforward. The electron beam current is set from zero to its nominal value at time $t  = 0~\mathrm{ns}$, with a rise time of $300~\mathrm{ps}$, while the initial radiation field is set to zero. With the given electron velocity, the time of flight through the $5\,\mathrm{cm}$ long photonic crystal slab is about $700~\mathrm{ps}$. The numerical calculations were performed in steps of about $1~\mathrm{fs}$ and extended over time intervals of up to $t  = 800~\mathrm{ns}$, which was sufficient for reaching a steady-state output in all investigated cases.

\begin{figure}[b]
	\includegraphics[width=0.75\linewidth]{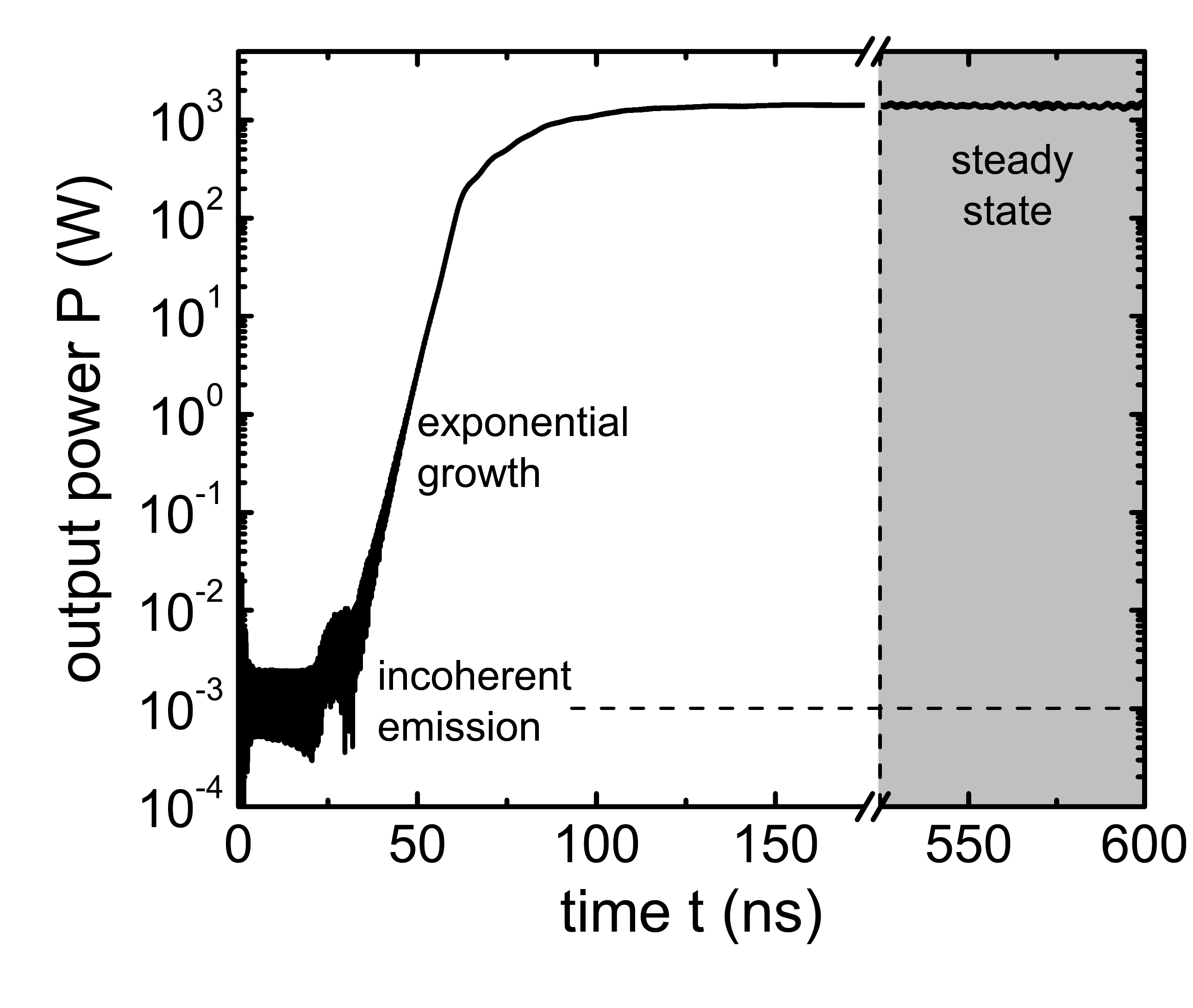}
	\caption{Temporal evolution of the generated output power when sending a continuous electron beam of $1\,\mathrm{A}$ through the crystal, turned on at $t=0$. Initially, the output consists of weak Cerenkov radiation (indicated by the horizontal dashed line). Thereafter, the output power increases exponential with time, before settling at a constant steady-state value (indicated by the gray area).
	 \label{figure3}}
\end{figure} 

A typical temporal evolution of the calculated output power is shown in Fig.~\ref{figure3}. The steady-state output is contained almost entirely in the fundamental transverse $\mathrm{TE}_{10}$ eigenmode of the waveguide. The field oscillates at a single frequency of $15.86~\mathrm{GHz}$, whereas other frequencies or modes of oscillation are suppressed by more than 2 orders of magnitude.

It can be seen that there is an initial temporal regime, which extends over the first $25~\mathrm{ns}$, where the output fluctuates and is rather weak, at the $1~\mathrm{mW}$-level. Comparing the initial output power with the kinetic power of the injected electron beam of  $I*E_e/e = 14.1~\mathrm{kW}$, where $E_e$ is the kinetic energy of a single electron, one finds an extremely low conversion efficiency, in the order of $10^{-7}$, which is typical for incoherent Cerenkov radiation \cite{Collins1938}. Based on the low efficiency, we address this initial regime to the incoherent superposition of single-electron responses \cite{Luo2003,Kremers2009a}. 
The intermediate regime in the dynamics, between $25~\mathrm{ns}$ and $60~\mathrm{ns}$, shows an exponential growth of power over many orders of magnitude (linear slope in the semi-logarithmic plot). Such growth is a clear signature of amplification by stimulated emission. After $60~\mathrm{ns}$ the power growth gradually diminishes.
After about $450~\mathrm{ns}$, in the final regime, the output is in steady state with a power of about $1.4~\mathrm{kW}$. This corresponds to a large conversion of about $10\%$ from the kinetic power of the electrons into radiation, which is six orders of magnitude higher than for the initial, incoherent emission. 

In order to investigate whether the strong growth in output power is accompanied with an increase in coherence as well, we calculated the degree of first-order temporal coherence, $g^{(1)}(\tau)$, where $\tau$ is the autocorrelation delay time \cite{Loudon1997}.  However, care has to be taken with a direct analysis of the fluctuating field in the initial temporal regime in Fig.~\ref{figure3}. The reason is that particle-in-cell calculations approximate electron beams via so-called quasi-particles (with an increased charge and mass compared to an electron), and that these are injected periodically with each numerical time step \cite{CST}. To exclude associated artifacts in the coherence function, we have applied an alternative method. The method makes use of the property of $g^{(1)}(\tau)$ that the emission from an ensemble of identical, uncorrelated emitters possesses the same coherence function as the emission from single emitters \cite{Loudon1997}. To obtain $g^{(1)}(\tau)$ of the single emitter in our case, we calculated the electromagnetic response to excitation with a single, ultra-short bunch of electrons. The duration and charge of the bunch was chosen sufficiently small, such that $g^{(1)}(\tau)$ was independent of these parameters ($<10~\mathrm{ps}$ and $<1~\mathrm{pC}$, respectively).

\begin{figure}[phtb]
       \includegraphics[width=0.80\linewidth]{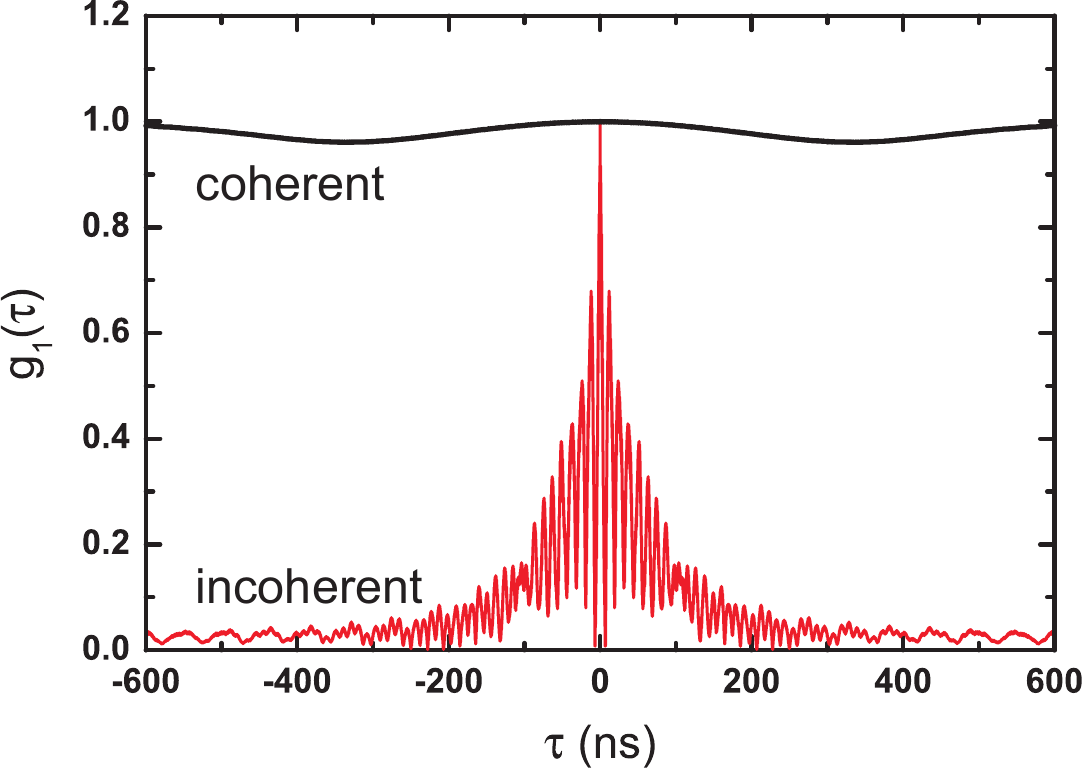}%
       \caption{First-order temporal coherence function, $g_{(1)}(\tau)$ of Cerenkov radiation from a photonic crystal slab. The red trace corresponds to the initial, weak emission in Fig.~\ref{figure3} and the black trace is calculated for steady-state, high-power emission. It can be seen that the coherence of the output radiation undergoes a transition from low coherence (coherence time 25 ns) towards high coherence.
       \label{figure4}}
\end{figure}

Figure~\ref{figure4} compares the calculated degree of first-order coherence for the initial regime of independently emitting electrons (red trace) with that for the final regime of steady-state emission (black curve). For the initial regime it can be seen that the coherence quickly decreases from its unity value (at $\tau=0$) when moving away from $\tau=0$. The corresponding coherence time, measured at $g^{(1)}(\tau)~=~0.5$, is rather short, about $25~\mathrm{ns}$ (HWHM), such that this regime can be named incoherent.

The black trace, in contrast, displays high values of $g^{(1)}(\tau)$ (from 0.96 to1.00) throughout the entire investigated range, with no noticeable signs of decay. This proves that the steady-state output possesses high coherence, as is typical for single-frequency laser oscillation. We address the residual variation \textit{vs.} $\tau$ to spurious oscillation at a neighbouring frequency, approximately 2 MHz off the main oscillation frequency.

\begin{figure}[tb]
	\includegraphics[width=0.70\linewidth]{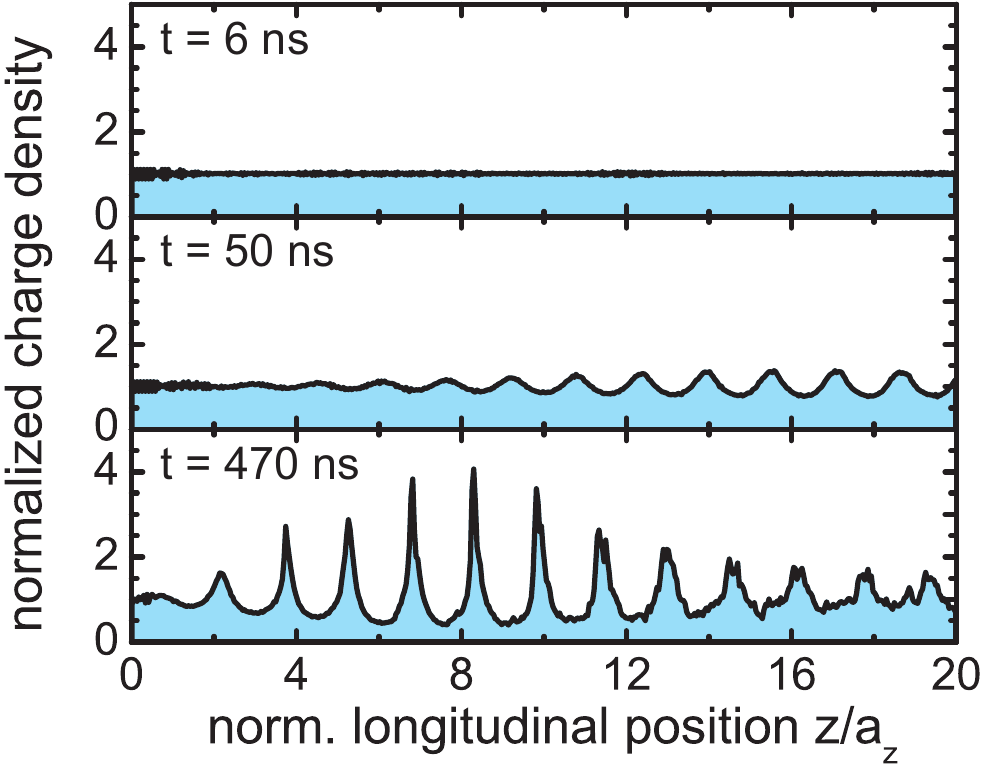}
	\caption{Line charge density of the electron beam (normalized to its value at $z = 0\,\mathrm{m}$) during incoherent emission ($t=6~$ns), during exponential growth ($t=50~$ns), and during coherent emission ($t=470~$ns). The \textit{z-}coordinate is normalized to the crystal period, $a_z$.
        \label{figure5}}
\end{figure}

The transition from initially incoherent emission to coherent emission in steady state becomes visible also in the transition from an initially homogeneous electron beam to a bunched beam, as described above. As an illustration, Fig.~\ref{figure5} presents the dynamical development of the line charge density of the electron beam along the longitudinal position inside the crystal. Results are shown for three representative times that correspond to the initial regime of incoherent emission ($t  = 6~\mathrm{ns}$), to the regime of exponential growth ($t  = 50~\mathrm{ns}$), and to the regime of steady-state coherent emission ($t  = 470~\mathrm{ns}$).
It can be seen that, during the initially weak and incoherent emission, the charge density is constant throughout the entire photonic crystal. During exponential growth, the electron beam develops bunching with an amplitude that grows towards the downstream end of the crystal. The period of the bunching is about $3.8~\mathrm{mm}$. This agrees well with the wavelength of $3.88~\mathrm{mm}$ of the velocity-matched spatial harmonic inside the photonic crystal as retrieved from Fig.~\ref{figure2} at $15.86~\mathrm{GHz}$ output frequency. In steady state, bunching is developed most strongly near the beginning of the crystal. Toward the end of the crystal, bunching becomes reduced again. The reduction can be addressed to a re-acceleration of electrons by the crystal-internal field. Re-acceleration absorbs radiation and thereby decreases (saturates) the net gain to a level that provides steady-state oscillation with constant output power.  

So far we have considered an ideal electron beam having zero energy spread and emittance. Increasing the relative energy spread $\frac{\delta\gamma}{\gamma}$ and normalized emittance $\varepsilon_{n}$ to more typical and experimentally realizable values, i.e., $\frac{\delta\gamma}{\gamma}\leq 2$~\% and $\varepsilon_{n} \leq 10$~mm~mrad, did not noticeably change the observed performance. 

\section{Conclusions}
In summary we have shown that a beam of free electrons traveling through a photonic crystal can generate powerful laser radiation with high coherence. This concept opens a wide avenue of novel possibilities in the field of laser physics. These include extension to three-dimensional photonic crystals, other crystal structures or materials (\textit{e.g.}, dielectrics or semiconductors), tuning of the output frequencies via the kinetic energy of the electron beam, or upscaling the output power with multiple electron beams. In principle, these novel possibilities can all be investigated using the capabilities of typical particle-in-cell simulation codes \cite{CST}. For instance, our initial calculations have shown that injecting several electron beams simultaneously, using an array of cathodes as indicated in Fig.~\ref{figure1}, increases the output power approximately in proportion with the number of beams. Alternatively, the same power can be obtained when distributing a given total current over a number of beams, each at a correspondingly lower current density. Such distributed pumping can serve to reduce undesired effects, such as increased Coulomb repulsion and losses, when scaling to higher frequencies via reducing the spatial period of the photonic crystal. We also found that the output frequencies can be tuned with the electron velocity, which is a result of velocity matching as in Fig.~\ref{figure2}. A rich field of phenomena can be explored and compared with standard lasers. Short pulses may be generated with temporal shaping of the electron current, or with a temporal chirp or modulation of the kinetic energy of the electrons. Wavefront steering and shaping may be obtained with a spatial chirp in the kinetic energy across multiple electron beams, and nonlinear conversion phenomena may be induced, \textit{e.g.}, the generation of harmonics of the fundamental laser frequency via phase velocity matching at additional frequencies. A fundamental property of our approach is that the output frequency is scalable to higher frequencies via the photonic crystal structure, enabling the generation of laser radiation in spectral ranges which are otherwise difficult to access.

\begin{acknowledgements}
The authors thank Ad Lagendijk for useful discussions. This research is supported by the Dutch Technology Foundation STW, which is part of the Netherlands Organisation for Scientific Research (NWO), and which is partly funded by the Ministry of Economic Affairs. 
\end{acknowledgements}

\bibliography{References}

\end{document}